# Spin Summations: A High-Performance Perspective

Paul Springer, AICES, RWTH Aachen
Devin Matthews, ICES, UT Austin
Paolo Bientinesi, AICES, RWTH Aachen

Besides tensor contractions, one of the most pronounced computational bottlenecks in the non-orthogonally spin-adapted forms of the quantum chemistry methods CCSDT and CCSDTQ, and their approximate forms—including CCSD(T) and CCSDT(Q)—are spin summations. At a first sight, spin summations are operations similar to tensor transpositions; a closer look instead reveals additional challenges to high-performance calculations, including temporal locality as well as scattered memory accesses. This publication explores a sequence of algorithmic solutions for spin summations, each exploiting individual properties of either the underlying hardware (e.g. caches, vectorization), or the problem itself (e.g. factorizability). The final algorithm combines the advantages of all the solutions, while avoiding their drawbacks; this algorithm, achieves high-performance through parallelization, vectorization, and by exploiting the temporal locality inherent to spin summations. Combined, these optimizations result in speedups between 2.4× and 5.5× over the NCC quantum chemistry software package. In addition to such a performance boost, our algorithm can perform the spin summations *in-place*, thus reducing the memory footprint by 2× over an *out-of-place* variant.

## 1. INTRODUCTION

In quantum chemistry, especially in high-accuracy calculations utilizing the popular coupled cluster family of methods [Čížek 1966; Bartlett and Musiał 2007; Helgaker et al. 2013], tensor algebra plays a central role and often accounts for the vast majority of the computational effort. The main tensor operation utilized is a tensor contraction, the multidimensional analogue of matrix multiplication, which is indeed usually expressed in terms of highly-optimized matrix multiplication kernels [Springer and Bientinesi 2016a; Matthews 2016].

Other essential tensor operations can consume a significant portion of the runtime, especially when they are poorly optimized relative to hand-tuned matrix multiplication kernels, or when they are practically bandwidth-bound. Tensor transpositions (permutations) [Wei and Mellor-Crummey 2014; Lyakh 2015; Springer et al. 2016;

Springer and Bientinesi 2016b] represent one of the bottlenecks, but recent work has shown that this cost can be minimized [Springer et al. 2017; Springer et al. 2016; Springer and Bientinesi 2016b] or even eliminated entirely [Matthews 2016; Springer and Bientinesi 2016a]. Instead, in high-accuracy calculations based on the efficient non-orthogonal spin-adaptation scheme [Čížek 1966; Kucharski and Bartlett 1992; Matthews et al. 2013], there remains an additional and significant bottleneck in the form of the "spin summation" operation [Matthews and Stanton 2015]. This operation is critical to forming the intermediates necessary to implement the fully factored form of the working equations, and also in the regularization of the tensor quantities to maintain numerical stability in the face of a linearly-dependent basis [Kucharski and Bartlett 1992; Adams and Paldus 1979]. The optimization of this important kernel is the focus of this work. In particular, we examine the set of spin summation operations summarized in Sec. 5.1, which cover all of the operations required for the non-orthogonally spin-adapted CCSDT [Noga and Bartlett 1987] and CCSDTQ [Kucharski and Bartlett 1992] quantum chemistry methods, as well as their approximate forms, such as CCSDT(Q) [Bomble et al. 2005].

Tensor transposition is related to spin summation in that the non-locality (in both space and time) of data references must be taken into account in order to achieve high efficiency on modern hardware. The primary difference between transposition and spin summation, though, is that spin summation entails data reuse of both the input and output tensors, while transposition accesses the tensors only once. For factorizable summations (defined below), there is also a difference between algorithms that perform optimal and sub-optimal numbers of floating-point operations. Thus, while specific optimization techniques from tensor transposition may be applied to the spin summation problem, a bottom-up analysis of algorithmic efficiency for this distinct class of problems is warranted.

The techniques and algorithms developed in this work are applicable to other important tensor operations beyond spin summations. A spin summation is essentially a linear sum of multiple tensor transpositions; this pattern appears in other tensor operations as well, for example in tensor (anti-)symmetrization [Hanrath and Engels-Putzka 2010; Lai et al. 2012], and in the aggregation of intermediate tensor products produced in distinct orderings [Baumgartner et al. 2005; Hartono et al. 2009]. Optimization of operations involving multiple transpositions is especially important for higher-dimensional tensors due to the proliferation of possible transpose variants.

To the authors' best knowledge, this work represents the first discussion of spin summations from a high-performance perspective. The contributions of this publications can be summarized as follows:

— We provide a detailed analysis of different algorithmic variants for the calculation of spin summations, exploiting various features of modern CPUs.
— We present a high-performance open-source implementation[1] for 3D and 4D spin summations used in computational chemistry, in particular in the non-orthogonally spin-adapted CCSDT and CCSDTQ methods and their various approximations. The algorithm is parallelized and vectorized, it exploits the CPU's cache hierarchy, and it reduces the amount of required floating-point operations while only using small buffers that fit into cache; together, these optimizations result in significant speedups (ranging from $2.4\times$ to $5.5\times$) over the reference NCC quantum chemistry software [Matthews and Stanton 2015], part of the CFOUR package [Stanton et al. 2014].

---

[1]The source code is publicly available at: `www.github.com/springer13/spin-summations`.

—We illustrate that the high-performance algorithm can also be used for *in-place* spin summations of 2D, 3D or 4D "square" tensors, reducing the memory footprint by $2\times$.

The remainder of this publication is structured as follows: Section 2 reviews related work on tensor transposition and related transformations such as unpacking of anti-symmetric tensors. Section 3 familiarizes the reader with spin summations, the nomenclature used, and the cache hierarchy of a modern CPU. Section 4 presents different implementations for spin summations and discusses their advantages and disadvantages. Section 5 introduces the exhaustive list of spin summations of interest and describes the measurement environment. Section 6 evaluates the performance of the different implementations and highlights their key features. Section 7 concludes our findings.

## 2. RELATED WORK

Matrix transposition (i.e. 2D tensor transposition) is a well studied operation, including optimizations for blocking, vectorization, unrolling, and software prefetching [McCalpin and Smotherman 1995; Mateescu et al. 2012; Lu et al. 2006; Goldbogen 1981; Vladimirov 2013; Chatterjee and Sen 2000]. For 3D tensors, the same optimizations are investigated in the context of out-of-place transpositions on CPUs [van Heel 1991; Jodra et al. 2015].

The optimization of arbitrary-dimensional tensor transpositions has gained more interest in recent years [Wei and Mellor-Crummey 2014; Lyakh 2015; Springer et al. 2016; Springer and Bientinesi 2016b]. Our previous work on the **T**ensor **T**ransposition **C**ompiler (TTC) [Springer et al. 2016; Springer and Bientinesi 2016b] relied on *code generation* to yield a nearly-optimal kernel for any given tensor transposition. While the generated code attained high performance in the general case, it was only applicable to tensor transpositions that were known at compile time. For this reason, we developed HPTT [Springer et al. 2017], a high-performance C++ library for tensor transpositions that maintains the desirable properties of TTC (e.g. explicit vectorization, loop reordering, autotuning) but avoids the code generation process.

While there has been some focus on in-place tensor transpositions [Ding 2001; He and Ding 2002; Catanzaro et al. 2014], their performance—in the general case—has been found to be limited.

In the context of data reuse in multidimensional operations, algorithms for unpacking anti-symmetric tensors into matrices were studied by Hanrath et al. [Hanrath and Engels-Putzka 2010]. They identify both "matrix-driven" and "tensor-driven" approaches, which are essentially equivalent to Algorithm 1 presented later. These approaches are also similar in terms of data access patterns to configuration interaction (CI) algorithms, which come in both "integral-driven" [Saunders and Lenthe 1983; Ansaloni et al. 2000; Klene and Robb 2000; Gan et al. 2003] and "configuration-driven" [Buenker and Krebs 1999; Jiang et al. 2009] varieties. In each of these cases, accesses to one multidimensional object is linearized at the expense of scattered accesses to another object. Blocking for the cache hierarchically has been studied and successfully applied in the context of stencil algorithms [Datta et al. 2008; Nguyen et al. 2010]; this is in several ways similar to the multidimensional tensor case.

## 3. BACKGROUND

To keep this publication self-contained, we introduce the used nomenclature and explain necessary architectural properties of a modern CPU.

### 3.1. Nomenclature

A $d$-dimensional tensor $\mathcal{A} \in \mathbb{R}^{n_1 \times ... \times n_d}$ is stored as a $d$-dimensional array adhering to the Fortran storage scheme (i.e. column-major). For instance, given a 3D tensor $\mathcal{A} \in \mathbb{R}^{n_1 \times n_2 \times n_3}$, the element $\mathcal{A}_{i_1 i_2 i_3}$, for any index $(i_1, i_2, i_3) \in \{0, ..., n_1 - 1\} \times \{0, ..., n_2 - 1\} \times \{0, ..., n_3 - 1\}$, is placed in the linearized memory location $\&(\mathcal{A}) + i_1 + i_2 n_1 + i_3 n_1 n_2$ ($\&$ denotes the "address of" operator). Due to the nature of our problem we are only concerned with three- and four-dimensional tensors for which the sizes of all dimensions are identical (i.e. $n_1 = n_2 = n_3 = n_4 = N$).

DEFINITION 1 (PERMUTATION). *Given a bijective function*
$$\phi : \{1, 2, ..., d\} \to \{1, 2, ..., d\}, \tag{1}$$
*we define the permutation $\pi$ that operates on $d$-tuples as:*
$$\pi : \mathbb{N}^d \to \mathbb{N}^d, \ (i_1, i_2, ..., i_d) \mapsto (i_{\phi(1)}, i_{\phi(2)}, ..., i_{\phi(d)}). \tag{2}$$
*Instead of defining $\pi$ explicitly, we use the shorter notation:* $\pi_{\phi(1)\phi(2)...\phi(d)}$.

***Example:*** $\pi_{321}$ applied to an index $(i_1, i_2, i_3)$ inverts its order and yields the permuted index $\pi_{321}(i_1, i_2, i_3) = (i_3, i_2, i_1)$.

DEFINITION 2 (TENSOR PERMUTATION). *Given a tensor $\mathcal{A} \in \mathbb{R}^{n_1 \times ... \times n_d}$ and a permutation $\pi$, we formally express a tensor permutation as*
$$\mathcal{B}_{i_1 i_2 ... i_d} \leftarrow \mathcal{A}_{\pi(i_1, i_2, ..., i_d)} \ \forall (i_1, i_2, ..., i_d) \in \{0, ..., n_1 - 1\} \times \{0, ..., n_2 - 1\} \times ... \times \{0, ..., n_d - 1\} \tag{3}$$
*Since this notation is quite verbose, we revert to a shorter form where $\pi$ acts as an operator on the entire tensor such that a tensor permutation can be written as $\mathcal{B} \leftarrow \pi(\mathcal{A})$. While we use the same symbol $\pi$ both as an operator that acts on indices as well as on tensors, these two case are easily distinguishable by the context.*

***Example:*** A matrix transposition of $\mathcal{A}, \mathcal{B} \in \mathbb{R}^{N \times N}$ can be expressed via the tensor permutation: $\mathcal{B} \leftarrow \pi_{21}(\mathcal{A})$.

DEFINITION 3 (SPIN SUMMATION). *In its most general form, a spin summation of a $d$-dimensional tensor $\mathcal{A}$ is a sum of $n$ tensor permutations $\omega_1, ..., \omega_n$ each scaled by some $\alpha_i \in \mathbb{R}$:*
$$\begin{aligned}\mathcal{B} &\leftarrow \sum_{i=1}^{n} \alpha_i \omega_i(\mathcal{A}) \\ &= (\alpha_1 \omega_1 + \alpha_2 \omega_2 + ... + \alpha_n \omega_n)(\mathcal{A}).\end{aligned} \tag{4}$$

***Example:*** $\mathcal{B} \leftarrow (\pi_{123} + \pi_{213} + \pi_{321} + \pi_{231} + \pi_{132} + \pi_{312})(\mathcal{A})$ is spin summation where all $\alpha_i$ are equal to one.

Most of the spin summations that we are interested in exhibit a very important property, which we refer to as *factorizable*.

DEFINITION 4 (FACTORIZABLE SPIN SUMMATION). *A spin summation $\mathcal{B} \leftarrow \sum_{i=1}^{n} \alpha_i \omega_i(\mathcal{A})$ is factorizable, if it can be separated into two successive spin summations such that:*
$$\mathcal{T} \leftarrow \sum_{i=1}^{\widehat{n}} \widehat{\alpha}_i \widehat{\omega}_i(\mathcal{A}), \tag{5}$$
$$\mathcal{B} \leftarrow \sum_{i=1}^{\widetilde{n}} \widetilde{\alpha}_i \widetilde{\omega}_i(\mathcal{T}), \tag{6}$$

with $\widehat{n} + \widetilde{n} \leq n$ and $\mathcal{T} \in \mathbb{R}^{n_1 \times \cdots \times n_d}$. *This definition can be applied recursively, factoring the initial spin summation into more than two stages.*

  ***Example:*** The spin summation

$$\mathcal{B} \leftarrow (\pi_{123} + \pi_{132} + \pi_{213} + \pi_{231} + \pi_{312} + \pi_{321})(\mathcal{A}) \tag{7}$$

can be factorized as

$$\mathcal{B} \leftarrow (\pi_{123} + \pi_{213})((\pi_{123} + \pi_{321} + \pi_{132})(\mathcal{A})). \tag{8}$$

Note that that every spin summation $\mathcal{B} \leftarrow \sum_{i=1}^{n} \alpha_i \omega_i(\mathcal{A})$ is factorizable without the constraint $\widehat{n} + \widetilde{n} \leq n$. For instance, let $\widehat{\omega}$ be an arbitrary permutation, then all factorizations of the following form are viable:

$$\mathcal{T} \leftarrow \widehat{\omega}(\mathcal{A}), \tag{9}$$

$$\mathcal{B} \leftarrow \sum_{i=1}^{n} \alpha_i \widetilde{\omega}_i(\mathcal{T}), \tag{10}$$

where $\widetilde{\omega}_i = (\omega_i \circ \widehat{\omega}^{-1})$ denotes the composition of $\omega_i$ and the inverse of $\widehat{\omega}$.

### 3.2. Cache Hierarchy

The cache hierarchy of modern CPUs bridges the ever widening gap between the performance of the CPU's floating-point units and its memory subsystem[2] by buffering frequently used data in fast *on-chip* memory (i.e. caches and registers).

  Caches exploit the fact that data, once loaded from main memory, is likely accessed again soon after; this effect is referred to as *temporal locality*. Thus, repeated accesses to the same memory location are served by the fast caches as opposed to the slow main memory. Another important feature of caches is that they fetch data at the granularity of a *cacheline* (typically 64 bytes = 16 single- = 8 double-precision elements). This means that an access to a memory location $l$ not only loads the element situated at $l$ but indeed its entire cacheline. Thus, a subsequent access to $l + 1$ (assuming that $l$ and $l + 1$ belong to the same cacheline) is served by the cache and does not require an additional load from main memory; this process is referred to as *spatial locality*.

  The caches of a CPU are organized into multiple levels, which vary in size, bandwidth and latency [Intel Corporation 2015]. The *Intel Xeon E5-2680 v3* CPU, for instance, has three levels: L1, L2, and L3 with decreasing speed and increasing size. While L1 and L2 are private to each core, the L3 (also known as last level cache, LLC) is shared between all the cores. Since modern CPUs use a *cache coherency protocol* [Patterson and Hennessy 2007; Bryant et al. 2003] to keep all the caches of a CPU coherent, it is important that different cores do not write to the same cacheline; such a situations is referred to as *false sharing* and results in *coherence traffic* among the cores that can significantly lower the performance.

  Modern *write-back* caches typically employ the *write-allocate* policy [Bryant et al. 2003]: A write-miss (i.e. a write to a memory location that is not in any level of the cache hierarchy) fetches the corresponding cacheline from main memory before updating it; this additional memory traffic is referred to as *write-allocate traffic*. Such a mechanism is favorable if the data is be accessed again in the near future. So called *non-temporal store* instructions avoid this *write-allocate traffic* by writing an entire

---

[2]The floating-point unit of one *Intel Xeon E5-2680 v3* core can issue two *fused-multiply-add*s ($\mathbf{c} \leftarrow \mathbf{a} \times \mathbf{b} + \mathbf{c}$) per cycle; each fused-multiply-add operates on $3 \times 8$ single-precision elements simultaneously. The full *Intel Xeon E5-2680 v3* CPU—assuming a turbo boost frequency of $3.1\,\mathrm{GHz}$—would require a staggering main memory bandwidth of $6652\,\mathrm{GiB/s}$ to keep all its floating-point units—across all its $12$ cores—busy. This value is in sharp contrast to the CPU's theoretical peak memory bandwidth of $63.3\,\mathrm{GiB/s}$.

---

**Algorithm 1:** Spatial locality in $\mathcal{B}$

---

1 **#pragma** omp parallel for collapse(2) schedule(static)
2 **for** $(i_3 = 0;\ i_3 < N;\ i_3\text{++})$ **do**
3    **for** $(i_2 = 0;\ i_2 < N;\ i_2\text{++})$ **do**
4       **for** $(i_1 = 0;\ i_1 < N;\ i_1\text{++})$ **do**
5          $\mathcal{B}_{i_1 i_2 i_3} \leftarrow \mathcal{A}_{i_1 i_2 i_3} + \mathcal{A}_{i_1 i_3 i_2} + \mathcal{A}_{i_2 i_1 i_3} + \mathcal{A}_{i_2 i_3 i_1} + \mathcal{A}_{i_3 i_1 i_2} + \mathcal{A}_{i_3 i_2 i_1}$

---

cacheline without fetching it from main memory prior to the write (see [Intel Corporation 2015] for more details).

### 3.3. Blocking

Any non-trivial real-world application has to utilize a CPU's rich cache hierarchy to attain good performance. One of the most powerful optimization techniques to enable such a design is *blocking*. *Blocking* refers to an reordering of the memory accesses to make better use of the fast on-chip memory (i.e. caches and registers). This optimization often entails a significant modification of the original algorithm. A very prominent operation that makes extensive use of this technique is matrix-matrix multiplication. The interested reader is referred to [Gunnels et al. 2001; Goto and Geijn 2008; Van Zee and van de Geijn 2015] for further details.

## 4. ROAD TO HIGH PERFORMANCE

The following discussion introduces several implementations of the spin summations in Eq. 7–8; this spin summation is small enough to serve as an example throughout this paper, yet it is rich enough to demonstrate all features of interest. The implementations presented in Sections 4.1–4.5 employ optimization strategies targeting different features of modern CPUs. We analyze their advantages and disadvantages and lay the foundation for a high-performance implementation (see Section 4.6) that combines the advantages of its predecessors while avoiding their drawbacks.

### 4.1. Algorithm 1: Locality in $\mathcal{B}$

Algorithm 1 is a direct translation of Eq. 7 into code. Provided that the compiler is smart enough to accumulate the six elements from $\mathcal{A}$ (Line 5) at the register level, this algorithm only requires one write to each memory location of $\mathcal{B}$, resulting in perfect *temporal locality* for $\mathcal{B}$. Moreover, the loops are ordered such that the output tensor $\mathcal{B}$ is traversed in a linear fashion. While this ordering leads to a preferable memory access pattern for $\mathcal{B}$, it causes an unfavorable access pattern for $\mathcal{A}$. Phrased differently, this algorithm perfectly exploits both the *spatial locality* as well as *temporal locality* in $\mathcal{B}$, but it only exhibits very limited *locality* in $\mathcal{A}$. More precisely, the *spatial locality* in $\mathcal{A}$ is limited to the accesses of $\mathcal{A}_{i_1 i_2 i_3}$ and $\mathcal{A}_{i_1 i_3 i_2}$ that have the same stride-one index (i.e. $i_1$) as the accesses to $\mathcal{B}$; all other accesses to $\mathcal{A}$ (i.e. $\mathcal{A}_{i_2 i_1 i_3}, \mathcal{A}_{i_2 i_3 i_1}, \mathcal{A}_{i_3 i_1 i_2}, \mathcal{A}_{i_3 i_2 i_1}$) have a different stride-one index. Likewise, *temporal locality* in $\mathcal{A}$ is only achieved if any of $i_1, i_2,$ or $i_3$ are identical; these situations are negligible. Furthermore, due to the lack of *spatial locality*, this implementation is not *vectorizable* without relying on expensive gather memory operations.

Thanks to the selected loop order, this algorithm can be efficiently parallelized via a single OpenMP statement (Line 1). This statement parallelizes all loops corresponding to the indices that are not the stride-one index of $\mathcal{B}$. This strategy is ideal in the sense that it avoids *false-sharing* between the threads while still providing as much parallelism as possible.

---

**Algorithm 2:** Temporal locality in $\mathcal{A}$ and $\mathcal{B}$

---

1 **#pragma** omp parallel for schedule(static,1)
2 **for** $(i_3 = 0;\ i_3 < N;\ i_3$++$)$ **do**
3    **for** $(i_2 = 0;\ i_2 \leq i_3;\ i_2$++$)$ **do**
4       **for** $(i_1 = 0;\ i_1 \leq i_2;\ i_1$++$)$ **do**
5          **for** $(\pi \in \{\pi_{123}, \pi_{132}, \pi_{213}, \pi_{231}, \pi_{312}, \pi_{321}\})$ **do**
6            $\mathcal{B}_{\pi(i_1,i_2,i_3)} \leftarrow \mathcal{A}_{\pi(i_1,i_2,i_3)} + \mathcal{A}_{\pi(i_2,i_1,i_3)} + \mathcal{A}_{\pi(i_3,i_2,i_1)} +$
7               $\mathcal{A}_{\pi(i_2,i_3,i_1)} + \mathcal{A}_{\pi(i_1,i_3,i_2)} + \mathcal{A}_{\pi(i_3,i_1,i_2)}$

---

### 4.2. Algorithm 2: Temporal Locality in $\mathcal{A}$ and $\mathcal{B}$

The key idea of Algorithm 2, outlined in this subsection, is to exploit the structure of spin summations and transform the scattered memory accesses to $\mathcal{A}$ into a structured form that makes use of the *temporal locality* in both the input and output tensor.

One noticeable difference to the previous algorithms is that this algorithm limits the range of the loops in Lines 2-4 such that the loop variables $i_1$, $i_2$, and $i_3$ do not span the entire iteration space $N^3$ but only a smaller tetrahedron ($0 \leq i_1 \leq i_2 \leq i_3 < N$).[3] The smaller iteration space is required due to the mechanics introduced by the new loop in Line 5. This loop iterates over all $d!$ permutations (Line 5) of the loop variables $i_1$, $i_2$, and $i_3$ (e.g. $(i_1,i_2,i_3)$, $(i_1,i_3,i_2)$, $(i_2,i_1,i_3)$).[4] The conceptual idea behind this loop is illustrated in Fig. 1. Consider the spin summation $\mathcal{B} \leftarrow (\pi_{123} + \pi_{321} + \pi_{132})(\mathcal{A})$ (see Fig. 1a): each element on the left (e.g. $\mathcal{B}_{i_1 i_2 i_3}$, $\mathcal{B}_{i_1 i_3 i_2}$) denotes a different entry of $\mathcal{B}$ for some fixed $i_1,i_2,i_3 \in \{0, 1, ..., N-1\}$ and is computed in one iteration of the innermost loop (see Algorithm 2, Line 5); likewise the elements on the right (e.g. $\mathcal{A}_{i_1 i_2 i_3}$, $\mathcal{A}_{i_1 i_3 i_2}$) denote different entries of $\mathcal{A}$. The connections between the nodes represent dependencies between these elements; the number of outgoing edges of each node depends on the actual spin summation (compare Fig. 1a and Fig. 1b). For instance, in Fig. 1a, the first iteration of the innermost loop computes $\mathcal{B}_{i_1 i_2 i_3} \leftarrow \mathcal{A}_{i_1 i_2 i_3} + \mathcal{A}_{i_3 i_2 i_1} + \mathcal{A}_{i_1 i_3 i_2}$, while the second iteration computes $\mathcal{B}_{i_1 i_3 i_2} \leftarrow \mathcal{A}_{i_1 i_3 i_2} + \mathcal{A}_{i_2 i_3 i_1} + \mathcal{A}_{i_1 i_2 i_3}$.

The colors in Fig. 1 identify different connected components;[5] this separation into connected components is not yet important, but we revisit this property in a Section 4.6.4.

Each iteration of the innermost loop accesses exactly six memory locations of $\mathcal{A}$ (Lines 6–7). Across all $d!$ iterations of this loop this totals $6 \times d!$ accesses to $\mathcal{A}$; however, only $d!$ of these memory locations are distinct. Thus, every element of $\mathcal{A}$, once loaded, is reused exactly six times, which results in perfect *temporal locality* for both tensors (as long as at least $d!$ elements can be stored in cache). Phrased differently, no element of $\mathcal{A}$ (or $\mathcal{B}$) needs to be loaded twice, if the memory subsystem would operate at the granularity of an element as opposed to a cacheline.

Figure 2 illustrates the underlying idea of Algorithm 2 on the two-dimensional example $\mathcal{B} \leftarrow (\alpha_1 \pi_{12} + \alpha_2 \pi_{21})(\mathcal{A})$. The two update statements (see Fig. 2, right) update distinct areas of $\mathcal{B}$ (denoted by the shaded regions). Both elements $\mathcal{B}_{i_1 i_2}$ (■) as well as $\mathcal{B}_{i_2 i_1}$ (■) depend on both $\mathcal{A}_{i_1 i_2}$ (■) and $\mathcal{A}_{i_2 i_1}$ (□). Thus, *temporal locality* is achieved

---

[3]We ignore special handling of cases where any of the indices $i_1$, $i_2$, and $i_3$ are identical for better readability. The actual implementation takes care of these cases.
[4]This loop is only visible in this listing, the actual implementation fully unrolls this loop such that all $d!$ updates statements are explicit, and omits redundant updates when any of $i_1$, $i_2$, or $i_3$ are equal.
[5]Two arbitrary nodes $a$ and $b$ belong to the same connected component, if and only if there exists a path from $a$ to $b$ or from $b$ to $a$.

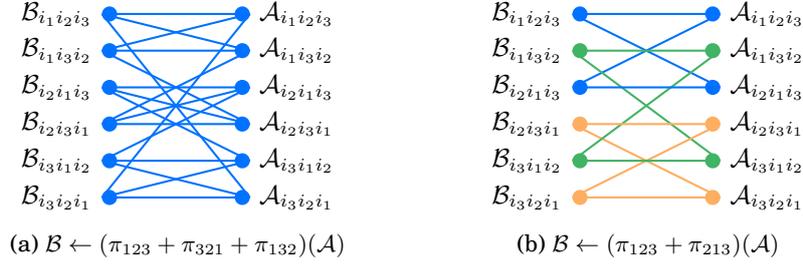

(a) $\mathcal{B} \leftarrow (\pi_{123} + \pi_{321} + \pi_{132})(\mathcal{A})$

(b) $\mathcal{B} \leftarrow (\pi_{123} + \pi_{213})(\mathcal{A})$

Fig. 1: Dependencies between elements of $\mathcal{B}$ and $\mathcal{A}$ for some fixed $i_1, i_2, i_3 \in \{0, 1, ..., N-1\}$ on the example of the two spin summations. Connected components are highlighted by different colors.

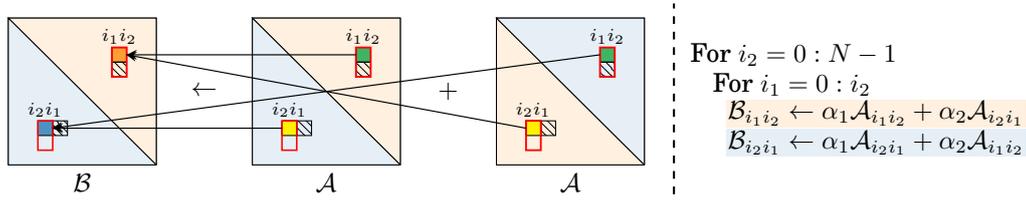

Fig. 2: Visualization of the 2D spin summation $\mathcal{B} \leftarrow (\alpha_1 \pi_{12} + \alpha_2 \pi_{21})(\mathcal{A})$. The shaded areas highlight the iteration space processed by the two update statements (right). Identical memory locations are identified by equally colored squares (i.e. both $\mathcal{A}$'s refer to the same memory location). Cachelines are denoted by red rectangles and next elements are denoted by striped patches.

since both update statements are executed back-to-back resulting in perfect reuse of the operands.

One of the fundamental problems of this algorithm, however, is its lack of *spatial locality*. As Fig. 2 highlights, the next elements (▨), accessed within the next iteration of the innermost loop, do not all belong to the previously-loaded cachelines (☐). In other words: This algorithm does not exhibit spatial locality for all memory accesses. While this problem does not seem too severe for the two-dimensional example, it becomes significantly accentuated in higher dimensions.

The missing *spatial locality* in $\mathcal{B}$ is especially troublesome for performance for three reasons: First and foremost, while this algorithm provides perfect *temporal locality* at the element level (i.e. no element needs to be loaded from main memory more than once), it does not exhibit this property at the cache line level (i.e. by the time another element—of a previously loaded—cachelines is accessed again this cacheline might already have been evicted from the caches). Second, the scattered memory accesses within the innermost loop (see Algorithm 2, Lines 6–7) prevent efficient *vectorization*. Finally, parallelizing any of the loops associated to $i_1$, $i_2$, and $i_3$ causes *false sharing* between the threads and thus hampers performance.

### 4.3. Algorithm 3: Reduced FLOP-count

So far we have targeted our optimizations at the caches. In contrast, this section aims to reduce the required amount of floating-point operations by exploiting the *factorizability* of spin summations (Definition 4). Algorithm 3 is implemented in terms of two dependent spin summations (Lines 1–5 and 6–10) that are directly translated from

---

**Algorithm 3:** Reduced FLOP-count
---
1 **#pragma** omp parallel for collapse(2) schedule(static)
2 **for** $(i_3 = 0; i_3 < N; i_3\text{++})$ **do**
3     **for** $(i_2 = 0; i_2 < N; i_2\text{++})$ **do**
4         **for** $(i_1 = 0; i_1 < N; i_1\text{++})$ **do**
5             $\mathcal{T}_{i_1 i_2 i_3} \leftarrow \mathcal{A}_{i_1 i_2 i_3} + \mathcal{A}_{i_3 i_2 i_1} + \mathcal{A}_{i_1 i_3 i_2}$
6 **#pragma** omp parallel for collapse(2) schedule(static)
7 **for** $(i_3 = 0; i_3 < N; i_3\text{++})$ **do**
8     **for** $(i_2 = 0; i_2 < N; i_2\text{++})$ **do**
9         **for** $(i_1 = 0; i_1 < N; i_1\text{++})$ **do**
10            $\mathcal{B}_{i_1 i_2 i_3} \leftarrow \mathcal{T}_{i_1 i_2 i_3} + \mathcal{T}_{i_2 i_1 i_3}$

---

their factorized form (Eq. 8); both of these implementations follow the design of Algorithm 1. Thus, the aforementioned drawbacks of this algorithm are also preserved.

We commence by observing that this algorithm requires a significant amount of *additional memory*, on the order of the problem size, to store the auxiliary tensor $\mathcal{T}$. Second, this algorithm reduces the memory accesses to $\mathcal{A}$ from $6N^3$ to $3N^3$, but it also adds three additional memory accesses (one write + two reads) per element of $\mathcal{T}$, which have not been required before. While the total amount of memory operations remained unchanged for this test case, other spin summations of interest significantly benefit from this optimization (see Section 6).

### 4.4. Algorithm 4: Reduced FLOP-count without Auxiliary Memory

---

**Algorithm 4:** Reduced FLOP-count + Fused
---
1 **#pragma** omp parallel for schedule(static,1)
2 **for** $(i_3 = 0; i_3 < N; i_3\text{++})$ **do**
3     **for** $(i_2 = 0; i_2 \leq i_3; i_2\text{++})$ **do**
4         **for** $(i_1 = 0; i_1 \leq i_2; i_1\text{++})$ **do**
5             **for** $(\pi \in \{\pi_{123}, \pi_{132}, \pi_{213}, \pi_{231}, \pi_{312}, \pi_{321}\})$ **do**
6                 $\mathcal{T}_{\pi(i_1 i_2 i_3)} \leftarrow \mathcal{A}_{\pi(i_1 i_2 i_3)} + \mathcal{A}_{\pi(i_3 i_2 i_1)} + \mathcal{A}_{\pi(i_1 i_3 i_2)}$
7             **for** $(\pi \in \{\pi_{123}, \pi_{132}, \pi_{213}, \pi_{231}, \pi_{312}, \pi_{321}\})$ **do**
8                 $\mathcal{B}_{\pi(i_1 i_2 i_3)} \leftarrow \mathcal{T}_{\pi(i_1 i_2 i_3)} + \mathcal{T}_{\pi(i_2 i_1 i_3)}$

---

Algorithm 4 is a combination of Algorithm 2 and Algorithm 3; its main goal is to combine their advantages while avoiding the need for an auxiliary tensor.

Instead of implementing each of the spin summations in Algorithm 4 in terms of Algorithm 1, we could alternatively use Algorithm 2. This change allows the loops in Lines 2–4 and 7–9 of Algorithm 3 to be fused, which—more importantly—reduces the size of the auxiliary tensor $\mathcal{T}$ from $N^d$ to merely $d!$ elements. Thus, accesses to $\mathcal{T}$ become negligible and do not need to be accounted for any longer; this also reflects positively on the number of memory operations that have to go to main memory. More precisely, if one neglects the presence of any cache and all other side-effects (e.g. *false sharing*), then one could come to the conclusion that Algorithm 4 is preferable over Algorithm 3 because it reduces the amount of memory operations from $(1 + 6)N^3$ to $(1 + 3)N^3$.

**Algorithm 5a:** Blocking

1 **#pragma** omp parallel
2 **#pragma** omp single
3 **for** ($\tilde{i}_3$ $0$; $\tilde{i}_3 \leq N$; $\tilde{i}_3$+ $bl$) **do**
4 | **for** ($\tilde{i}_2$ $0$; $\tilde{i}_2 \leq \tilde{i}_3$; $\tilde{i}_2$+ $bl$) **do**
5 | | **for** ($\tilde{i}_1$ $0$; $\tilde{i}_1 \leq \tilde{i}_2$; $\tilde{i}_1$+ $bl$) **do**
6 | | | **#pragma** omp task
7 | | | macroKernel($\mathcal{A}$, $\mathcal{B}$, $\tilde{i}_1, \tilde{i}_2, \tilde{i}_3$)

**Algorithm 5b:** Macro Kernel

1 **Function** macroKernel($\mathcal{A}$, $\mathcal{B}$, $\tilde{i}_1, \tilde{i}_2, \tilde{i}_3$):
2 | **for** ($i_3$ $\tilde{i}_3$; $i_3 < \tilde{i}_3 + bl$; $i_3$++) **do**
3 | | **for** ($i_2$ $\tilde{i}_2$; $i_2 < \tilde{i}_2 + bl$; $i_2$++) **do**
4 | | | **for** ($i_1$ $\tilde{i}_1$; $i_1 < \tilde{i}_1 + bl$; $i_1$++) **do**
5 | | | | **for** ($\pi \in \{\pi_{123}, \pi_{132}, \pi_{213}, \pi_{231}, \pi_{312}, \pi_{321}\}$) **do**
6 | | | | | $\mathcal{T}_{\pi(i_1 i_2 i_3)} \leftarrow \mathcal{A}_{\pi(i_1 i_2 i_3)} + \mathcal{A}_{\pi(i_3 i_2 i_1)} + \mathcal{A}_{\pi(i_1 i_3 i_2)}$
7 | **for** ($i_3$ $\tilde{i}_3$; $i_3 < \tilde{i}_3 + bl$; $i_3$++) **do**
8 | | **for** ($i_2$ $\tilde{i}_2$; $i_2 < \tilde{i}_2 + bl$; $i_2$++) **do**
9 | | | **for** ($i_1$ $\tilde{i}_1$; $i_1 < \tilde{i}_1 + bl$; $i_1$++) **do**
10 | | | | **for** ($\pi \in \{\pi_{123}, \pi_{132}, \pi_{213}, \pi_{231}, \pi_{312}, \pi_{321}\}$) **do**
11 | | | | | $\mathcal{B}_{\pi(i_1 i_2 i_3)} \leftarrow \mathcal{T}_{\pi(i_1 i_2 i_3)} + \mathcal{T}_{\pi(i_2 i_1 i_3)}$

**Algorithm 5:** The blocked algorithm (left) is implemented in terms of a *macro-kernel* (right). Each invocation of the macro-kernel calculates $d!$ blocks of size $bl^3$ of the output tensor. $\mathcal{T}$ refers to auxiliary memory. We ignore special handling of cases where any of $\tilde{i}_1, \tilde{i}_2,$ or $\tilde{i}_3$ are identical to improve the readability of the pseudo code.

While this algorithm reduces the FLOP-count and removes the requirement for an auxiliary tensor, the drawbacks mentioned in Section 4.2 still apply.

### 4.5. Algorithm 5: Blocking

Algorithm 5 presented in this section capitalizes on the analysis of the previous sections and combines the benefits of Algorithms 1–4. It forms the foundation for the high-performance implementation presented in Section 4.6.

While Algorithm 4 already expresses some desirable properties such as *reduced FLOP-count* as well as *temporal locality* for individual elements, this *locality* does not extend to entire cachelines. Thus, the main objective of Algorithm 5 is to provide *temporal locality* at the cache level (i.e. a cacheline is only loaded from main memory once). The key optimization to enable such a design is *blocking* (see Section 3.3).

Algorithm 5 is very similar to Algorithm 4 with the critical difference that Algorithm 5 operates on blocks of size $bl^3$ as opposed to scalars. The calculation of these blocks is separated out into a so called *macro-kernel* (see Algorithm 5b). All invocations of the macro-kernel are independent from one another, and can therefor be computed in parallel by different threads. In contrast to the loop-based parallelism (Algorithms 1–4), this algorithm uses tasked-based parallelism. More precisely, a single thread (Algorithm 5a, Line 2) generates all the tasks (Algorithm 5a, Lines 6–7), which are then dynamically processed by any of the available threads.

The most important parameter in this algorithm is the blocksize $bl$.[6] This parameter is subject to several constraints imposed by our high performance objective: First, $bl$ should be a multiple of the cacheline size—in elements (e.g. 8 for double-precision calculations)—to optimally exploit *spatial locality*. Second, all $2 \times d!$ blocks—per thread—must fit into the shared last level cache. Moreover, $bl$ should be as large as possible to facilitate more memory accesses to consecutive elements and (if possible) take advantage of the *adjacent cacheline prefetching* feature of modern CPUs; more-

---

[6]For the course of this discussion, we assume that $N$ is evenly divisible by the blocksize $bl$. However, the actual implementation does not have this constraint and generalizes to non-square blocks and, thus, works for arbitrary sizes.

---

**Algorithm 6:** High-Performance macro-kernel. Identical subtensors are highlighted by the same color. The notation $\mathcal{T}_{\tilde{i}_1 \tilde{i}_2 \tilde{i}_3}[i_1, i_2, i_3]$ is equivalent to $\mathcal{T}_{\tilde{i}_1+i_1, \tilde{i}_2+i_2, \tilde{i}_3+i_3}$; it is merely used to separate the loop variables, $i_1, i_2, i_3$ from the offsets, $\tilde{i}_1, \tilde{i}_2, \tilde{i}_3$.

1  **Function** macroKernel($\mathcal{A}$, $\mathcal{B}$, $\tilde{i}_1$, $\tilde{i}_2$, $\tilde{i}_3$):
2     **for** ($i_2 = 0$; $i_2 < bl$; $i_2$++) **do**
3        **for** ($i_3 = 0$; $i_3 < bl$; $i_3$++) **do**
4           **for** ($i_1 = 0$; $i_1 < bl$; $i_1$++) **do**
5              $\mathcal{T}_{\tilde{i}_1 \tilde{i}_2 \tilde{i}_3}[i_1,i_2,i_3] \leftarrow \mathcal{A}_{\tilde{i}_1 \tilde{i}_2 \tilde{i}_3}[i_1,i_2,i_3] + \mathcal{A}_{\tilde{i}_3 \tilde{i}_2 \tilde{i}_1}[i_3,i_2,i_1] + \mathcal{A}_{\tilde{i}_1 \tilde{i}_3 \tilde{i}_2}[i_1,i_3,i_2]$
6              $\mathcal{T}_{\tilde{i}_1 \tilde{i}_3 \tilde{i}_2}[i_1,i_2,i_3] \leftarrow \mathcal{A}_{\tilde{i}_1 \tilde{i}_3 \tilde{i}_2}[i_1,i_2,i_3] + \mathcal{A}_{\tilde{i}_2 \tilde{i}_3 \tilde{i}_1}[i_3,i_2,i_1] + \mathcal{A}_{\tilde{i}_1 \tilde{i}_2 \tilde{i}_3}[i_1,i_3,i_2]$
7              $\mathcal{T}_{\tilde{i}_2 \tilde{i}_1 \tilde{i}_3}[i_1,i_2,i_3] \leftarrow \mathcal{A}_{\tilde{i}_2 \tilde{i}_1 \tilde{i}_3}[i_1,i_2,i_3] + \mathcal{A}_{\tilde{i}_3 \tilde{i}_1 \tilde{i}_2}[i_3,i_2,i_1] + \mathcal{A}_{\tilde{i}_2 \tilde{i}_3 \tilde{i}_1}[i_1,i_3,i_2]$
8              $\mathcal{T}_{\tilde{i}_2 \tilde{i}_3 \tilde{i}_1}[i_1,i_2,i_3] \leftarrow \mathcal{A}_{\tilde{i}_2 \tilde{i}_3 \tilde{i}_1}[i_1,i_2,i_3] + \mathcal{A}_{\tilde{i}_1 \tilde{i}_3 \tilde{i}_2}[i_3,i_2,i_1] + \mathcal{A}_{\tilde{i}_2 \tilde{i}_1 \tilde{i}_3}[i_1,i_3,i_2]$
9              $\mathcal{T}_{\tilde{i}_3 \tilde{i}_1 \tilde{i}_2}[i_1,i_2,i_3] \leftarrow \mathcal{A}_{\tilde{i}_3 \tilde{i}_1 \tilde{i}_2}[i_1,i_2,i_3] + \mathcal{A}_{\tilde{i}_2 \tilde{i}_1 \tilde{i}_3}[i_3,i_2,i_1] + \mathcal{A}_{\tilde{i}_3 \tilde{i}_2 \tilde{i}_1}[i_1,i_3,i_2]$
10             $\mathcal{T}_{\tilde{i}_3 \tilde{i}_2 \tilde{i}_1}[i_1,i_2,i_3] \leftarrow \mathcal{A}_{\tilde{i}_3 \tilde{i}_2 \tilde{i}_1}[i_1,i_2,i_3] + \mathcal{A}_{\tilde{i}_1 \tilde{i}_2 \tilde{i}_3}[i_3,i_2,i_1] + \mathcal{A}_{\tilde{i}_3 \tilde{i}_1 \tilde{i}_2}[i_1,i_3,i_2]$
   // second part $\mathcal{B}_{\pi(i_1 i_2 i_3)} \leftarrow \mathcal{T}_{\pi(i_1 i_2 i_3)} + \mathcal{T}_{\pi(i_2 i_1 i_3)}$ omitted ...

---

over, larger blocks reduce the parallelization overhead introduced by the tasked-based parallelization scheme. These constraints respectively suggest $bl = 16$ and $bl = 8$ for 3D and 4D spin summations. More precisely, 3D and 4D spin summations respectively require $384$ KiB and $1536$ KiB of cache per thread; this fits well into the available last level cache of modern CPUs (e.g. an *Intel Xeon E5-2680 v3* offers $2560$ KiB LLC per core).

The scratchpad memory $\mathcal{T}$ (see Algorithm 5b) is allocated as a contiguous chunk of memory in order to avoid severe cache conflicts (caused by the limited *associativity* of the caches). Moreover, each thread initializes its portion of $\mathcal{T}$ individually to account for the *non-uniform memory accesses* of modern CPUs. We point out that—despite our best efforts to keep all accessed blocks in cache—this algorithm may still experience few redundant loads from main memory due to the accesses to $\mathcal{A}$;[7] we suggest an optimization for this scenario in Section 4.6.4.

The small—yet important—distinctions between operating on blocks (Algorithm 5) as opposed to scalars (Algorithm 4) entails several positive effects on performance: First and foremost, provided that the blocks remain in some level of the cache hierarchy, this is the first algorithm that achieves *temporal locality* for entire cachelines for both $\mathcal{A}$ and $\mathcal{B}$. Second, this algorithm does not experience any *false-sharing* between the threads since $bl$ is chosen to be a multiple of the cacheline size. Finally, the abundantly available parallelism and the dynamic tasked-based scheduling together provide good *load-balancing*; for instance, a tensor as small as $40^4$ elements (equivalent to $19.5$ MiB) already offers $70$ independent tasks.

All in all, Algorithm 5 exhibits many desirable properties, but it does not offer *spatial locality* in the sense that a cacheline—once loaded into L1 cache—is consumed at once; most of the elements of this cacheline are accessed at a later stage. This delayed access necessitates an additional load of the cacheline from some higher level (i.e. L2 or L3) of the cache hierarchy. This disadvantage is addressed in the next section.

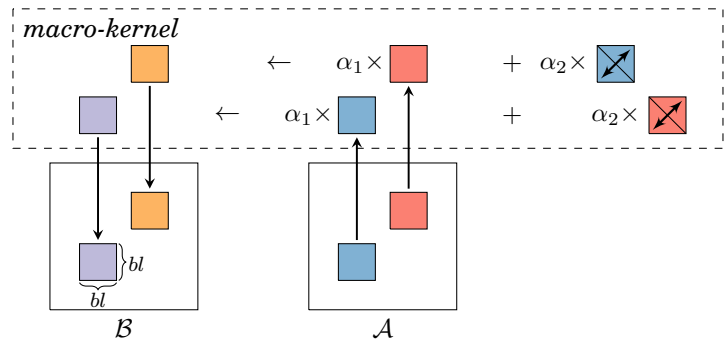

Fig. 3: Schematic overview for the vectorization of the 2D spin summation $\mathcal{B} \leftarrow (\alpha_1 \pi_{12} + \alpha_2 \pi_{21})(\mathcal{A})$. Same blocks are colored identically; each block denotes a 2D face consisting of $bl^2$ elements. The transpose of a block is indicated by an double-edged arrow.

### 4.6. Algorithm 6: High Performance

We combine the underlying ideas of the previous implementations into a high-performance algorithm. All optimizations introduced in this section only affect the macro-kernel (Algorithm 5b); the remainder is identical to Algorithm 5a.

*4.6.1. Restructured Block-traversal.* The main challenge for a high-performance macro-kernel is the strided memory accesses pattern of Algorithm 5b. For instance, each of the loop variables (i.e. $i_1$, $i_2$, $i_3$) associated to the loops in Lines 2–4 of Algorithm 5b is a stride-one index for either the output or the input tensor at some iteration of the loop in Line 5. This circumstance poses significant problems with respect to *spatial locality* and makes efficient vectorization impossible.

Algorithm 6 outlines the new macro-kernel; despite the fact that it looks quite different from its predecessor (Algorithm 5b), they are indeed logically identical. The critical difference, however, is that Algorithm 6 traverses the $d!$ blocks (each of size $bl^d$) in a different order; identical blocks—those blocks for which $\tilde{i}_1$, $\tilde{i}_2$, and $\tilde{i}_3$ appear in the same order—are colored identically. Algorithm 6 can be deduced from Algorithm 5b via the following steps: (1) Pull the loop starts (i.e. $\tilde{i}_1, \tilde{i}_2$ and $\tilde{i}_3$) into the $\mathcal{A}_{...}$ notation; this simple change renders all loops indistinguishable from one another (i.e. all loops have the same start position and loop trip count). (2) Now that the loops in Lines 2–4 are indistinguishable, rename the loop variables within each update statement individually (Lines 5–10); different names correspond to different traversal orders. While any traversal is valid from a correctness perspective, some traversal orders are preferable from a high-performance point of view. Thus, we rename the loop variables in each line such that both the input and output tensors are only accessed by **at most** two different stride-one indices (i.e. the leftmost indices: $i_1$ and $i_3$).

Now that we only have to deal with at most two different stride-one indices (Algorithm 6), we can reorder the loops the loops (Lines 2–4) such that the loops associated to the stride-one indices are the innermost loops. This strategy greatly improves *spatial locality* and enables efficient *vectorization* along these loops (see next Section).

*4.6.2. Vectorization.* The vectorization scheme for spin summations is quite similar to that of tensor transpositions [Springer et al. 2016; Springer and Bientinesi 2016b] in the sense that both operate on 2D faces of arbitrary-dimensional tensors. However, in contrast to tensor transpositions, spin summations have to load multiple faces and add them together—not just one. All these faces have to share at most two different stride-one indices to utilize fully-vectorized memory operations (avoiding scatter and gather).

---
[7]The blocks of $\mathcal{A}$ are not consecutive in main memory and, therefor, may cause cache conflicts.

Figure 3 illustrates the vectorization on the example of the 2D spin summation $\mathcal{B} \leftarrow (\alpha_1 \pi_{12} + \alpha_2 \pi_{21})(\mathcal{A})$. The vectorized macro-kernel proceeds as follows: (1) Load the two $bl^2$ blocks of $\mathcal{A}$ (■, ■) into the cache (mostly L1 and L2); (2) add the corresponding transposed blocks (■, ■) to yield the output (■, ■); finally, (3) store these blocks to the correct locations in $\mathcal{B}$. All these steps are fully vectorized (see [Springer et al. 2016; Springer and Bientinesi 2016b] for details), exploiting *spatial locality*.

Given that our vectorization requires no more than two different stride-one indices, the natural question is whether we can *always* restructure the traversal within the macro-kernel such that this requirement is met. While most of the spin summations of interest are vectorizable without any additional work, some require an additional *regularization step*. For instance, the spin summation $\mathcal{B} \leftarrow (\alpha_1 \pi_{123} + \alpha_2 \pi_{213} + \alpha_3 \pi_{321})(\mathcal{A})$ has three different stride-one indices; thus, any renaming strategy of the indices is bound to fail. However, such a situation can be avoided by a *regularization step* that transposes the input tensor $\mathcal{A}$ prior to the actual spin summation and account for this change at a later stage (see Eq. 9–10). More precisely, instead of computing $\mathcal{B} \leftarrow (\alpha_1 \pi_{123} + \alpha_2 \pi_{213} + \alpha_3 \pi_{321})(\mathcal{A})$ directly, one could also transpose $\mathcal{T} \leftarrow \pi_{213}(\mathcal{A})$ and compute the output as $\mathcal{B} \leftarrow (\alpha_1 \pi_{213} + \alpha_2 \pi_{123} + \alpha_3 \pi_{231})(\mathcal{T})$ that only has two different stride-one indices; this procedure is applicable to all the spin summations that we are interested in and enables efficient vectorization.

*4.6.3. Non-Temporal Stores.* To avoid the *write-allocate traffic* (see Section 3.2) incurred by the writes to the output tensor $\mathcal{B}$, we use *non-temporal store* instructions whenever possible. It is important to realize that non-temporal stores should only be applied to the writes to $\mathcal{B}$ and not to $\mathcal{T}$, since the latter is reused repeatedly. More importantly, failing to adhere to this policy results in poor performance because cachelines written via *non-temporal stores* are marked as invalid, and a later access to such a cacheline incurs a redundant read from main memory.

To make efficient use of *non-temporal stores*, it is critical to reorder the loops of the macro-kernel such that writes to the same cacheline occur soon after one another.[8] Such writes enables the underlying microarchitecture to *write-combine* successive writes to the same cacheline into a single memory transaction. Thus, we order the loops within the macro-kernel such that the loop associated to the stride-one index of $\mathcal{B}$ is the innermost loop (see Algorithm 6); failing to do so results in redundant reads/writes from/to main memory and, thus, significantly lowers the performance.

*4.6.4. Cache Optimization.* Section 4.5 already pointed out that the blocking scheme might require redundant reads from main memory if the accesses to $\mathcal{A}$ cause *conflict misses* [Bryant et al. 2003]. This problem may be averted by (1) larger caches, (2) larger cache associativity or (3) smaller cache requirements of the algorithm. While we clearly cannot do anything about (1) and (2), there is some opportunity for (3).

Recall that the blocking scheme (see Section 4.5) requires each thread to keep all its blocks—traversed within the macro-kernel—in some level of the cache hierarchy. More precisely, computing a spin summation over $d$-dimensional double-precision tensors with $m$ threads and block of size $bl^d$ requires a minimal L3 cache size $C^{\min}$

$$C^{\min} = bl^d \times m \times x \times 8 \, \text{bytes}, \tag{11}$$

---

[8]On an AVX-enabled architecture one needs two *non-temporal stores* (each writing 32 byte) to update an entire cacheline (64 byte).

where $x = 2 \times d!$ represents the number of blocks traversed within the macro-kernel. To reduce $C^{\min}$ one may try to decrease either $bl$, $m$, or $x$. As discussed previously, decreasing $bl$ is not desirable and leads to poor performance (data not shown). Similarly, measurements show that reducing the number of threads $m$ to a point where some CPU cores are idle also yields bad performance (data not shown) because the benefits of more last level cache—per thread—are countered by the reduced parallelism. This leaves us with the last option: reducing $x$.

Consider the spin summation $\mathcal{B} \leftarrow (\pi_{123} + \pi_{213})(\mathcal{A})$ outlined in Fig. 1b, where identically colored nodes correspond to update statements that belong to the same connected component and therefore share operands. For instance, the two update statements $\mathcal{B}_{i_1 i_2 i_3} \leftarrow \mathcal{A}_{i_1 i_2 i_3} + \mathcal{A}_{i_2 i_1 i_3}$ and $\mathcal{B}_{i_2 i_1 i_3} \leftarrow \mathcal{A}_{i_2 i_1 i_3} + \mathcal{A}_{i_1 i_2 i_3}$—highlighted in blue—share the same elements (i.e. $\mathcal{B}_{i_1 i_2 i_3}, \mathcal{B}_{i_2 i_1 i_3}, \mathcal{A}_{i_1 i_2 i_3}, \mathcal{A}_{i_2 i_1 i_3}$) but they do not use any of the elements belonging to other updates; the same analysis extends to blocks instead of elements. Thus, blocks of $\mathcal{B}$ that belong to different connected components may be computed separately from one another without losing any *spatial locality* or *temporal locality*. For instance, separating a spin summation over $d$-dimensional tensors into $c$ connected components reduces the amount of blocks $x$ from $2 \times d!$ to $2 \times \frac{d!}{c}$. Thus, this technique can reduce the pressure on the cache hierarchy significantly.

This optimization is not applicable to all spin summations: For instance, for the spin summations $\mathcal{B} \leftarrow (\pi_{123} + \pi_{321} + \pi_{132})(\mathcal{A})$ (Fig. 1a), a decomposition into independent connected components is not possible.

*4.6.5. In-Place.* *In-place* spin summations (i.e. $\mathcal{A} \leftarrow \sum_{i=1}^{n} \alpha_i \omega_i(\mathcal{A})$) facilitate another effective opportunity to avoid the *write-allocate traffic*. While high-performance implementations of *in-place* tensor transpositions for 2D, non-square tensors are a hard problem [Ding 2001; He and Ding 2002], our algorithm is (almost) immediately applicable to *in-place*, "hyper-square" spin summations. The support for *in-place* spin summations is mostly contributed to the auxiliary scratchpad memory $\mathcal{T}$ and the *temporal locality* of the algorithm. For instance, instead of directly computing the factorized spin summation $\mathcal{T} \leftarrow (\pi_{123} + \pi_{321} + \pi_{132})(\mathcal{A})$ followed by $\mathcal{B} \leftarrow (\pi_{123} + \pi_{213})(\mathcal{T})$, we change the latter statement to $\mathcal{A} \leftarrow (\pi_{123} + \pi_{213})(\mathcal{T})$. This is possible since all memory locations in $\mathcal{A}$ that have already been used once and are not used again. Some spin summations of interest, however, do not require any scratchpad memory because they are not *factorizable* (e.g. $\mathcal{B} \leftarrow (\pi_{123} + \pi_{321} + \pi_{132})(\mathcal{A})$). For these spin summations, we introduce an additional preprocessing step—similar to the regularization step outlined in Section 4.6.2—which copies the blocks of $\mathcal{A}$ into temporal blocks $\mathcal{T}$. For instance, the in-place spin summation $\mathcal{A} \leftarrow (\pi_{123} + \pi_{321} + \pi_{132})(\mathcal{A})$ is computed in two stages: (1) $\mathcal{T} \leftarrow \pi_{123}(\mathcal{A})$ followed by (2) $\mathcal{A} \leftarrow (\pi_{123} + \pi_{321} + \pi_{132})(\mathcal{T})$; the amount of additional memory is very limited, because the size of these blocks is chosen such that they all fit into the caches simultaneously.

*4.6.6. Summary.* Table I summarizes the properties of all previously mentioned algorithms. While Algorithms 1–5 score positively on some properties and negatively on others, the high-performance algorithm (see Section 4.6) is the only one that scores positively across all properties.

## 5. MEASUREMENT ENVIRONMENT & BENCHMARK

We evaluate the performance of Algorithms 1–6 on a two-socket system consisting of two *Intel Xeon E5-2680 v3* CPUs (12 cores each). The compiler of choice is Intel's *icpc 16.0.2* using the `-O3 -qopenmp -xHost` options. Moreover, we use one thread per phys-ical core (i.e. 24 threads in total) and pin logically neighboring threads to physically neighboring cores.[9]

| Algo | TMP locality $\mathcal{A}$ | $\mathcal{B}$ | SP locality $\mathcal{A}$ | $\mathcal{B}$ | VEC | LB | FLOP | MEM | IP | NTS |
|---|---|---|---|---|---|---|---|---|---|---|
| 1 | - | + | - | + | - | + | - | + | - | - |
| 2 | + | + | - | - | - | - | - | + | - | - |
| 3 | - | + | - | + | - | + | + | - | - | - |
| 4 | + | + | - | - | - | - | + | + | - | - |
| 5 | + | + | o | o | - | + | + | + | + | - |
| 6 | + | + | + | + | + | + | + | + | + | + |

Table I: Summary over all features of each implementations. *TMP locality*, *SP locality*, *VEC, LB, FLOP, MEM, IP, NTS* respectively stand for temporal locality, spatial locality, vectorization, load balancing, reduced FLOP-count, reduced auxiliary memory requirements, support for in-place summations, and support for non-temporal stores. '+', 'o' and '-' respectively reflect positively, neutrally or negatively on the corresponding property.

Given a spin summation (see Eq. 4) summing over $n$ permutations of an input tensor $\mathcal{A} \in \mathbb{R}^{N^d}$, we report the bandwidth BW and the floating-point performance FP as follows:

$$\text{BW} = \frac{2 \times N^d \times \texttt{sizeof(double)}}{2^{30} \times \text{Time}} \text{ GiB/s,} \tag{12}$$

$$\text{FP} = \frac{2 \times n \times N^d}{10^9 \times \text{Time}} \text{ GFLOP/s.} \tag{13}$$

Notice that the bandwidth metric is especially conservative since we only account for one read of the input tensor $\mathcal{A}$, independent of the number of permutations $n$. The floating-point performance, on the other hand, accounts for the *effective* FLOP-count corresponding to the FLOPS performed by Algorithms 1 and 2.

To put our results into perspective, we respectively report the STREAM [McCalpin 1995] *copy*- ($\mathbf{b} \leftarrow \mathbf{a}$), *scale*- ($\mathbf{b} \leftarrow \alpha\mathbf{a}$), *add*- ($\mathbf{c} \leftarrow \mathbf{a}+\mathbf{b}$) and *triad*-bandwidth ($\mathbf{c} \leftarrow \alpha\mathbf{a}+\mathbf{b}$) for the full system: 101.2 GiB/s, 101.3 GiB/s, 107.6 GiB/s and 107.7 GiB/s. Moreover, the total theoretical peak floating-point performance of our two-socket system is 1113.6 GFLOP/s; MKL attains 845 GFLOP/s for a large matrix-matrix multiplication.

All performance results are based on the minimum execution time over several runs to get the most stable timings; the caches are cleared before every run (i.e. all measurements start on cold data). All computations use double-precision floating-point accuracy.

### 5.1. Benchmark

Table II lists all spin summations required for the non-orthogonally spin-adapted CCSDT [Noga and Bartlett 1987] and CCSDTQ [Kucharski and Bartlett 1992] quantum chemistry methods, as well as their approximate forms such as CCSDT(Q) [Bomble et al. 2005]. We test the performance of our implementations with three different problem sizes: *small*, *medium*, and *large* respectively corresponding to tensors roughly of size 70 MiB, 320 MiB, and 1200 MiB. Unless otherwise mentioned, we present results for the *medium*-sized problems.

| ID | Test Case |
|----|-----------|
| 1  | $\mathcal{B} \leftarrow (2 - \pi_{213})((2 - \pi_{321} - \pi_{132})(\mathcal{A}))$ |
| 2  | $\mathcal{B} \leftarrow (2 - \pi_{321} - \pi_{132})(\mathcal{A})$ |
| 3  | $\mathcal{B} \leftarrow (2 - \pi_{213} - \pi_{132})(\mathcal{A})$ |
| 4  | $\mathcal{B} \leftarrow (2 - \pi_{213} - \pi_{321})(\mathcal{A})$ |
| 5  | $\mathcal{B} \leftarrow (2-\pi_{2134})((2-\pi_{3214}-\pi_{1324})((2-\pi_{4231}-\pi_{1432}-\pi_{1243})(\mathcal{A})))$ |
| 6  | $\mathcal{B} \leftarrow (2 - \pi_{3214} - \pi_{1324})((2 - \pi_{4231} - \pi_{1432} - \pi_{1243})(\mathcal{A}))$ |
| 7  | $\mathcal{B} \leftarrow (2 - \pi_{2134} - \pi_{1324})((2 - \pi_{4231} - \pi_{1432} - \pi_{1243})(\mathcal{A}))$ |
| 8  | $\mathcal{B} \leftarrow (2 - \pi_{2134} - \pi_{3214})((2 - \pi_{4231} - \pi_{1432} - \pi_{4231})(\mathcal{A}))$ |
| 9  | $\mathcal{B} \leftarrow (2 - \pi_{4231} - \pi_{1432})((2 - \pi_{3214} - \pi_{1324} - \pi_{4231})(\mathcal{A}))$ |
| 10 | $\mathcal{B} \leftarrow (2 - \pi_{2134} - \pi_{1432})((2 - \pi_{3214} - \pi_{1324} - \pi_{4231})(\mathcal{A}))$ |
| 11 | $\mathcal{B} \leftarrow (2 - \pi_{2134} - \pi_{4231})((2 - \pi_{3214} - \pi_{1324} - \pi_{1234})(\mathcal{A}))$ |
| 12 | $\mathcal{B} \leftarrow (2 - \pi_{4231} - \pi_{1243})((2 - \pi_{2134} - \pi_{1324} - \pi_{1432})(\mathcal{A}))$ |
| 13 | $\mathcal{B} \leftarrow (2 - \pi_{3214} - \pi_{1243})((2 - \pi_{2134} - \pi_{1324} - \pi_{1432})(\mathcal{A}))$ |
| 14 | $\mathcal{B} \leftarrow (2 - \pi_{3214} - \pi_{4231})((2 - \pi_{2134} - \pi_{1324} - \pi_{1432})(\mathcal{A}))$ |
| 15 | $\mathcal{B} \leftarrow (2 - \pi_{1432} - \pi_{1243})((2 - \pi_{2134} - \pi_{3214} - \pi_{4231})(\mathcal{A}))$ |
| 16 | $\mathcal{B} \leftarrow (2 - \pi_{1432} - \pi_{1243})((2 - \pi_{2134} - \pi_{3214} - \pi_{4231})(\mathcal{A}))$ |
| 17 | $\mathcal{B} \leftarrow (2 - \pi_{1432} - \pi_{1432})((2 - \pi_{2134} - \pi_{3214} - \pi_{4231})(\mathcal{A}))$ |
| 18 | $\mathcal{B} \leftarrow (2 - \pi_{4231} - \pi_{1432} - \pi_{1243})(\mathcal{A})$ |
| 19 | $\mathcal{B} \leftarrow (2 - \pi_{3214} - \pi_{1324} - \pi_{1243})(\mathcal{A})$ |
| 20 | $\mathcal{B} \leftarrow (2 - \pi_{2134} - \pi_{1324} - \pi_{1432})(\mathcal{A})$ |
| 21 | $\mathcal{B} \leftarrow (2 - \pi_{2134} - \pi_{3214} - \pi_{4231})(\mathcal{A})$ |

Table II: Spin summations. Depending on the context, "2" is short for $2\pi_{123}$ or $2\pi_{1234}$.

## 6. PERFORMANCE EVALUATION

Figure 4 summarizes the performance of the discussed algorithms across the set of 21 spin summations (see Table II). We make the following key observations: (1) Algorithms 1–4 (■, ■, ■, ■) show ambivalent performance for the different test cases with no clear loser nor winner among them. For instance, Algorithm 1 performs well on test cases 2, 3, 4, 19, and 20 but it greatly lags behind any of the Algorithms 2–4 for test cases 1, 5–17. (2) While Algorithm 3 (■) is implemented in terms of Algorithm 1 (■), they exhibit significantly different performance results for all test cases for which the FLOP-count can be reduced (i.e. 1, 5–17). For instance, the difference between the attained bandwidth of Algorithms 1 and 3 is largest for test case 5; this matches our expectations since this test case corresponds to the spin summation for which the FLOP-count can be reduced the most. (3) Algorithm 5 (■) is consistently faster than any of its predecessors; its superior performance can be contributed mostly to its improved *temporal locality*. (4) Algorithm 6 (■), with all its optimizations (e.g. *vectorization*, *regularization*), yields a significant improvement over its blocked, but *non-vectorized*, predecessor (■) across all spin summations. (5) While test cases 6–17 require the same amount of FLOPS and memory accesses, they attain noticeably different bandwidth. This is a similar observation that we have already made in a previous publication [Springer et al. 2016], indicating that some transpositions are inherently more difficult than others. For instance, test cases 4, 8, 11, 14–17, and 21 require an additional regularization step (see Section 4.6.1) to enable vectorization. (6) The attained effective floating-point performance (see Fig. 4b) greatly varies across the benchmark, reflecting that some spin summations benefit from the FLOP-count optimization more than others. More precisely, the theoretical FLOP-count reduction for test cases 1, 2–4, 5, 6–17, and 18–21 respectively is $2\times$, $1\times$, $4\times$, $2\times$, and $1\times$.

---

[9]The *thread affinity* is set via the environment variable KMP_AFFINITY = compact,1.

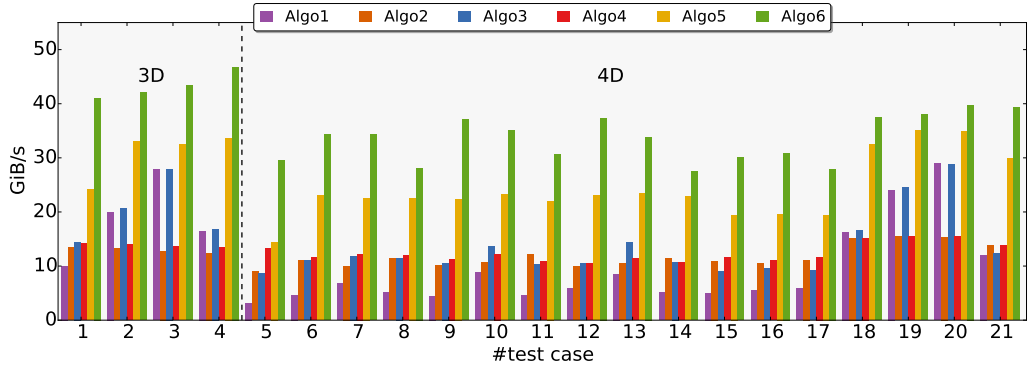

(a) Bandwidth.

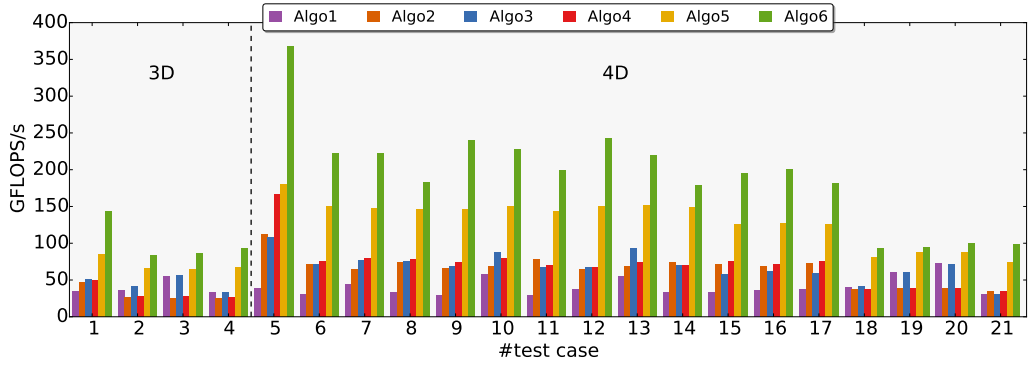

(b) Effective floating-point performance.

Fig. 4: Performance of all algorithms across the benchmark.

### 6.1. Optimizations

Figure 5 highlights the effect of various optimizations (see Sections 4.6.3–4.6.5) on the performance of Algorithm 6. The reference for the speedups is Algorithm 6 with all optimizations deactivated; similarly, the measurements labeled *NTS, cache-opt* (◆) correspond to the results in Fig. 4.

It is evident from Fig. 5 that: First, the cache optimization (▽, see Section 4.6.4) only has an effect on the performance for 4D spin summations; this is expected because 3D spin summations only require a small portion of the available last level cache, making cache conflicts less likely. Moreover, the positive effect of the cache optimization is evident both in the presence or absence of *non-temporal* stores. Second, activating *non-temporal stores* (▲, see Section 4.6.3) is critical for performance. The speedup due to *non-temporal stores*, in some case, can be as high as $1.5\times$; such a speedup is perfectly in line with the fact that the amount of data transfered from the caches is reduced by 33% (see Section 4.6.3).[10] Finally, *in-place* spin summations (●, see Section 4.6.5) do not only require less memory, they also increase the performance over its out-of-place counter part (◆) by up to $1.2\times$.

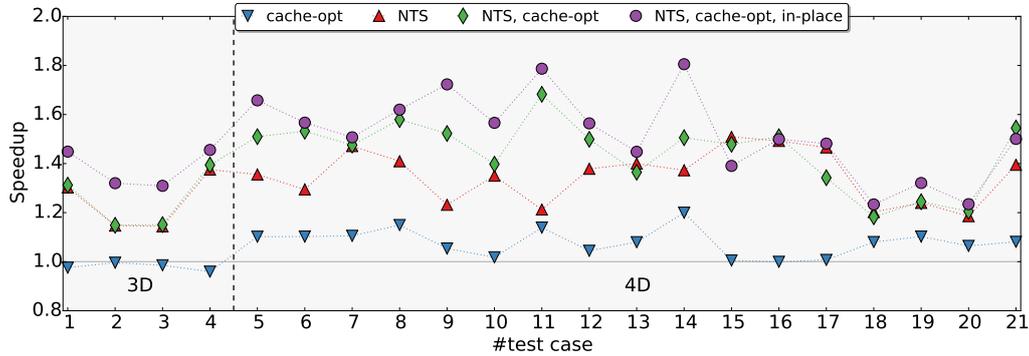

Fig. 5: Impact of various optimization techniques on the performance of Algorithm 6. Reference: Algorithm 6 without any optimizations. ▽ cache-optimization; ▲ *non-temporal stores*; ◆ *non-temporal stores* and cache-optimization; ● all optimizations and *in-place*.

### 6.2. Summary

Figure 6 summarizes the performance of Algorithm 6 for different problem sizes. Since Algorithm 2 constitutes the current implementation of spin summation in the NCC quantum chemistry software package [Matthews and Stanton 2015], Fig. 6b reports the speedup of Algorithm 6 with respect to Algorithm 2.

The average speedups for the *small* (▽), *medium* (▲), and *large* (◆) problems respectively vary between $2.4 - 4.3\times$, $2.4 - 3.8\times$, and $3.3 - 5.5\times$. Thus, our high-performance implementation substantially outperforms the reference for all spin summations and tested problem sizes. While the speedups for the *large* problems are the highest, we point out that these speedups are caused by the combination of (1) slightly increased performance of Algorithm 6 (see Fig. 6a), and (2) decreased performance of Algorithm 2 (i.e. the reference).

### 7. CONCLUSION

We tackled the problem of making the computation of 3D and 4D spin summations as efficient as possible. We presented a systematic way to restructure the memory accesses so that both the *temporal* and *spatial locality*, inherent to the operation, are exploited; the resulting algorithm also takes advantage of the *factorizablity* of spin summations, thus significantly reducing the required FLOP-count. The lesson learned is that it is not one optimization in isolation that makes the difference, but rather the combination of all of them. For instance, the main techniques introduced in Section 4.1–4.4 (e.g. *spatial locality, temporal locality*, and reduced FLOP-count) show ambiguous performance results if applied in isolation. However, the integration of all these ideas into Algorithm 6, yields noticeable speedups, between $2.4\times$ and $5.5\times$ over the current reference implementation. Furthermore, our algorithm allows for spin summations to be performed *in-place*; this desirable property not only reduces the memory footprint by a factor of two, but also reflects positively on the overall performance.

---

[10]The maximum theoretical speedup of $1.5\times$ assumes that the accessed data to both tensors is entirely cached and that the write-allocate traffic is avoided.

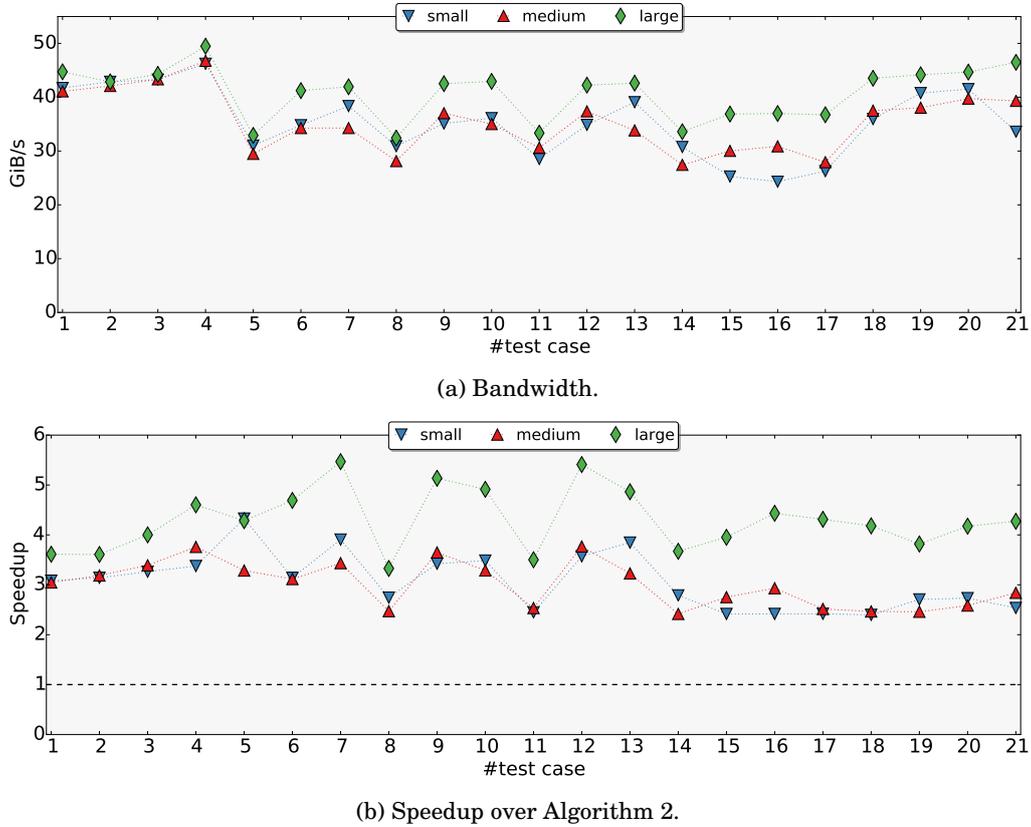

(a) Bandwidth.

(b) Speedup over Algorithm 2.

Fig. 6: Performance of Algorithm 6 for three different problem sizes: small (70 MiB), medium (320 MiB), large (1200 MiB).

We plan to incorporate our high-performance implementation of spin summation based on Algorithm 6 into the NCC quantum chemistry software package. In concert with other ongoing work in optimizing tensor contraction [Matthews 2016; Springer and Bientinesi 2016a], the speedups achieved in this work should allow for large-scale CCSDT, CCSDTQ, and CCSDT(Q) calculations to run at near-peak efficiency on modern multi- and many-core systems. This is exciting, given that previous experience has been that these calculations only achieved a small fraction of peak performance.

As already mentioned, the sums of tensor transpositions appears in other operations beyond spin summations. We believe the techniques detailed in this work are also applicable for tensor unpacking, and for the anti-symmetrization operations required in open-shell calculations using an unrestricted Hartree-Fock reference.

**ACKNOWLEDGMENT**

Financial support from the Deutsche Forschungsgemeinschaft (DFG) through grant GSC 111, and from the Intel Corporation through Parallel Computing Center grants to RWTH Aachen and UT Austin is gratefully acknowledged. Devin A. Matthews is an Arnold O. Beckman Postdoctoral Fellow and gratefully acknowl-edges support from the Arnold and Mabel Beckman Foundation. Furthermore, we thank our colleagues in the HPAC research group for many fruitful discussions and valuable feedback.